\documentclass{icrc29}
\usepackage{graphicx}
\usepackage{amssymb,amsmath,times}
\begin{document}
\title[A search for neutrinos from the galactic plane with AMANDA-II ...]{A
  Search for High-energy Muon Neutrinos from the Galactic Plane with AMANDA-II}
\author[J.L. Kelley et al.] {J.L. Kelley$^a$ for the IceCube Collaboration$^b$ \\
  (a) Physics Department, University of Wisconsin, Madison, WI 53706 USA \\
  (b) For a full author list, see arXiv:astro-ph/0509330
}
\presenter{}

\maketitle


\begin{abstract}

Interactions of cosmic rays with the galactic interstellar medium produce
high-energy neutrinos through the decay of charged pions and kaons.  We
report on a search with the AMANDA-II detector for muon neutrinos from the
region of the galactic plane below the horizon from the South Pole
($33^{\circ} <$ galactic longitude $< 213^{\circ}$).  Data from 2000 to
2003 were used for the search, representing a total of 807 days of livetime
and 3329 candidate muon neutrino events.  No excess of events was
observed. For a spectrum of $E^{-2.7}$ and Gaussian spatial distribution
($\sigma = 2.1^\circ$) around the galactic equator, we calculate a flux
limit of $4.8\times
10^{-4}\ \mathrm{GeV}^{-1}\ \mathrm{cm}^{-2}\ \mathrm{s}^{-1}\ \mathrm{sr}^{-1}$ in 
the energy range from 0.2 to 40 TeV.

\end{abstract}


\section{Introduction}
\label{Introduction}

High-energy neutrinos are produced in the disk of the Galaxy as cosmic rays
interact with the interstellar medium (ISM), creating charged pions and
kaons.  Because of the low density of the ISM, the particles produced
typically decay before interacting again, and the energy spectrum of the
neutrinos follows the primary cosmic ray spectrum of $E^{-2.7}$.  Most
models of this emission predict a flux that is proportional to the column
density of the ISM, and thus highest towards the Galactic Center
\cite{Berezinsky93}, \cite{Ingelman96}.

The AMANDA-II detector, a subdetector of the IceCube experiment, is an
array of 677 optical modules buried in the ice at the geographic South Pole
which detects the \v{C}erenkov radiation from charged particles produced in
neutrino interactions with matter \cite{Andres01}.  In particular, muons
produced in charged-current $\nu_{\mu}$ and $\bar{\nu}_{\mu}$ interactions
deposit light in the detector with a track-like topology, allowing us
to use directional reconstruction to reject the large background of
down-going atmospheric muon events.  Up-going atmospheric neutrinos are the
primary remaining background for this search.  Because we restrict
ourselves to events originating below the horizon, we are not sensitive to
the region near the Galactic Center; rather, we perform a search for
neutrinos from the outer region of the galactic plane, $33^\circ < $ galactic
longitude $< 213^\circ$.  Using the parametrization
in ref. \cite{Ingelman96} with an average ISM column density in this region
of $0.8\times10^{22}\ \mathrm{cm^{-2}}$, we expect at Earth an average
$\nu_{\mu} + \bar{\nu}_{\mu}$ flux of $3.9 \times
10^{-6}\ \mathrm{GeV}^{-1}\ \mathrm{cm}^{-2}\ \mathrm{s}^{-1}\ \mathrm{sr}^{-1}$.


\section{Signal Hypothesis and Simulation}

The actual distribution of the ISM in the galactic plane is quite irregular
\cite{Nakanishi03}, so we use a simplified signal hypothesis.  Because the
ISM column density in the outer Galaxy does not vary too much (see e.g. the
map by Bloemen in ref. \cite{Berezinsky93}), we model the neutrino flux as
isotropic in galactic longitude.  The expected profile in galactic latitude
has not been discussed in detail in the literature, although we can study
models of $\gamma$-ray emission as a guide.  A recent model by Strong
\textit{et al.}  of the $\gamma$-ray emission from $\pi^0$ decay in a
somewhat lower energy range (4-10 GeV) has an approximately Gaussian
profile with $\sigma \approx 2.1^\circ$ around the galactic equator
\cite{Strong04}.

For our initial signal hypothesis, we have simulated a line source from the
galactic equator that is isotropic in galactic longitude.  As discussed
later, we also use two other spatial profiles: a diffuse flux near the
galactic equator, and a Gaussian with $\sigma = 2.1^\circ$.  The spectral
slope is assumed to be -2.7, but other values ranging from -2.4 to -2.9 are
also tested (for specific models, see e.g. \cite{Strong04}).  We do not model
the change of slope at the knee of the cosmic ray spectrum, since the
resulting difference in number of events is negligible.  Possible point
sources in the galactic plane have also not been considered.  To produce
the signal Monte Carlo (MC), we use a reweighting method to transform an
isotropic distribution of simulated events \cite{Ahrens03} to a line source
originating from the galactic equator.  The absolute normalization of the
simulated signal flux is adjusted after normalizing the atmospheric neutrino
MC to the data sample.


\section{Data Sample}

The data sample used for this analysis consists of 3329 candidate muon
neutrino events collected from 2000 to 2003, representing 807 days of
livetime.  The event selection involved a number of quality criteria to
reject mis-reconstructed down-going muon events, and was optimized for a
broad sensitivity to an $E^{-2}$ to $E^{-3}$ spectrum.  This sample was
originally used for a point-source neutrino search, and details of the data
selection procedure are presented elsewhere in these proceedings
\cite{Ackermann05}.  During the design and optimization of event selection
criteria, the right ascension of the data events was scrambled in
accordance with our blind analysis procedures. 


\section{Background Estimation and Optimization of Selection Criteria}

\begin{figure}[ht]
\centering
\includegraphics[scale=0.42]{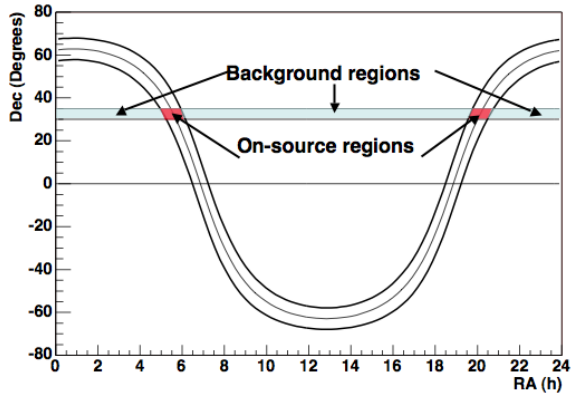}
\caption{Regions of the sky used for on-source event counting
  and background estimation for a particular declination slice.}
\label{fig_signal_bg}
\end{figure}

Because of our isotropic detector response in right ascension, we can use the
data to estimate the background in a point-source search by counting
events in a declination band around the sky.  For this analysis, however,
the source is extended across a large range of declinations, requiring a
modification of this technique.  We define the \textit{on-source region} as
the band of sky within $B$ degrees of the galactic equator.  The on-source
region is first divided into slices of equal declination $5^\circ$ wide, and
the background is estimated by counting the number of events in the
declination band outside the on-source region and scaling by the solid
angle ratio (see fig. \ref{fig_signal_bg}).  The total number of on-source
and background events is then calculated by summing over the declination bands.

The direction of the events is the most useful parameter in distinguishing a
galactic signal, and we optimize our sensitivity by varying the size of the
on-source region.  The region size is chosen to minimize the model
rejection potential \cite{Hill03}, the ratio of the average event upper
limit at the 90\% confidence level to the number of expected signal events
at a given reference flux.

\begin{figure}[ht]
\centering
\begin{minipage}[t]{7.2cm}
\includegraphics[width=0.9\textwidth]{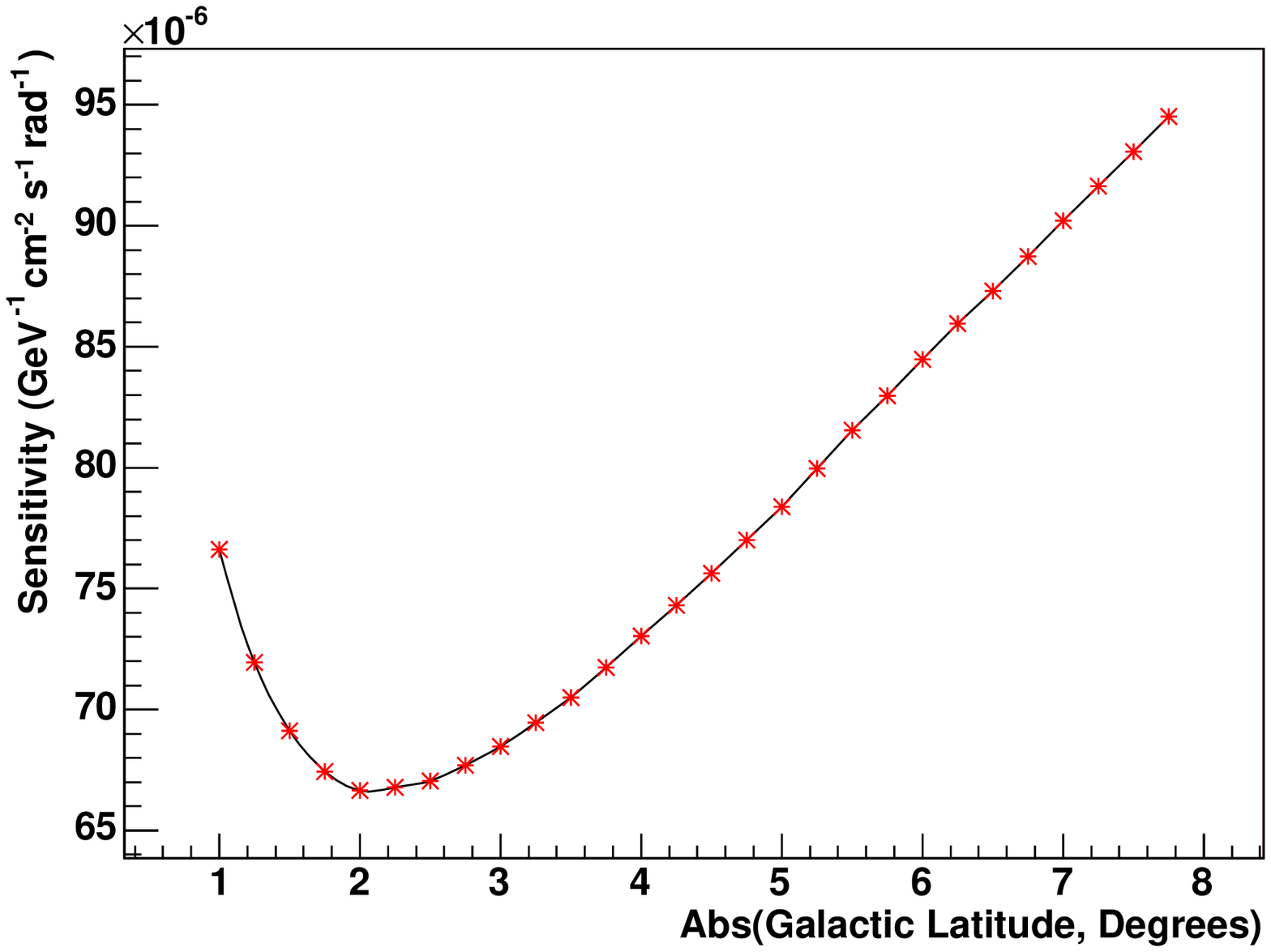}
\caption{Sensitivity to a line source as a function of on-source region size.}
\label{fig_sens_line}
\end{minipage}
\hfill
\begin{minipage}[t]{7.2cm}
\includegraphics[width=0.9\textwidth]{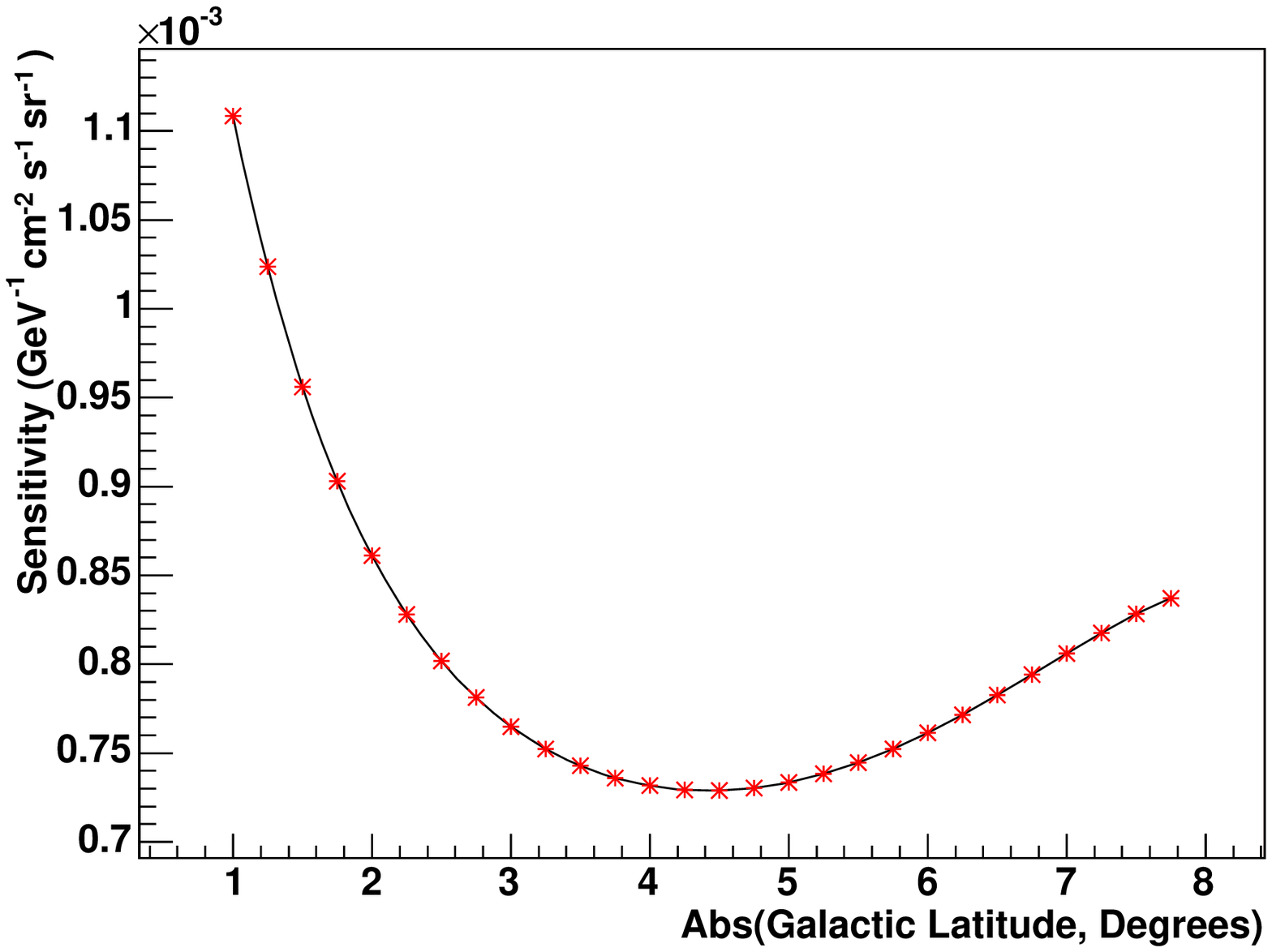}
\caption{Sensitivity to a Gaussian source ($\sigma = 2.1^\circ$) as a
  function of on-source region size.}
\label{fig_sens_gauss}
\end{minipage}
\hfill
\end{figure}

The optimal on-source region size for a line source was found to be
$B=2^\circ$ (see fig. \ref{fig_sens_line}).  This optimization is, however,
a bit artificial since it is primarily determined by our line spread
function.  To obtain a value for more realistic flux distributions, we use
an analytical method to estimate the sensitivity to a diffuse flux in the
on-source region, as well as to a Gaussian signal profile of a given width.
Our sensitivity to another signal profile is the flux level at which the
total number of events in the on-source region is equal to that of the
original line-source flux.  This approximation is valid as long as the
zenith angle does not vary too much over the on-source region.

First, we convert the line-flux sensitivity $\Phi_{line}$ (angular units of
$\mathrm{rad}^{-1}$) to a diffuse-flux sensitivity $\Phi_{diff}$ (angular
units of $\mathrm{sr}^{-1}$).  We integrate the line flux over $\pi$
radians of galactic longitude, divide by the solid angle $\Omega_{gal}$ of
the on-source region, and include an efficiency factor $\eta$ as the
fraction of signal events in that region:

\begin{equation}
\label{sens_diff}
\Phi_{diff} = \eta\ \pi\ \Phi_{line}\ /\ \Omega_{gal}\ .
\end{equation}

A similar procedure can be used to estimate the sensitivity to a Gaussian
signal profile.  The convolution of a Gaussian signal of width $\sigma_{sig}$ with
the line spread function (also approximated as a Gaussian, of width
$\sigma_{lsf} \approx 1.5^\circ$) results in a wider Gaussian.  As before,
by integrating to equalize the number of events in the angular region, we
solve for the Gaussian peak sensitivity $\Phi_{peak}$ in terms of the line
source sensitivity $\Phi_{line}$:

\begin{equation}
\label{sens_gauss}
\Phi_{peak}(B) = \frac{\Phi_{line}(B)}{\sqrt{2\pi(\sigma_{lsf}^2 +
    \sigma_{sig}^2)}} \ \operatorname{erf}(B / \sqrt{2}\sigma_{lsf})
    \ /\ \operatorname{erf}(B / \sqrt{2 (\sigma_{lsf}^2 +
      \sigma_{sig}^2)})\ .
\end{equation}

Using the relationship between the line-flux region size $B$ and the
sensitivity $\Phi_{line}$, we can reoptimize for the wider Gaussian signal
profile.  For a Gaussian with $\sigma_{sig} = 2.1^\circ$, we find an
optimal on-source region of $B=4.4^\circ$ (see fig. \ref{fig_sens_gauss}).

\section{Results}

After unblinding the data, we observe no excess of events.  We first
calculate a limit on the line source flux at the 90\% confidence level, and
then calculate corresponding limits for diffuse and Gaussian spatial profiles
using the analytical expressions above (eqns. \ref{sens_diff} 
and \ref{sens_gauss}).  These results are presented in table
\ref{tbl_results} for a signal spectrum of $E^{-2.7}$.  The energy range of
these limits, incorporating the central 90\% of the signal spectrum after
all selection criteria, is 0.2 to 40 TeV.  The calculation has been
repeated using spectral slopes from -2.4 to -2.9, resulting in
diffuse-flux limits ranging from $5.3\times10^{-5}$ to
$3.1\times10^{-3}\ \mathrm{GeV}^{-1} \mathrm{cm}^{-2} \mathrm{s}^{-1}
\mathrm{sr}^{-1}$.

Because the signal flux is normalized using atmospheric neutrino MC, the
largest systematic error is the uncertainty of $\sim$30\% on the absolute
atmospheric neutrino flux, and this has been incorporated into the limits
\cite{Conrad03},\cite{Hill03b}.   Possible unquantified sources of error are 
variations in the width of the Gaussian signal profile, and the offset of
the peak flux from the galactic equator.

\begin{table}[ht]
\begin{center}
\begin{tabular}{|c|c|c|c|c|c|c|c|} 
\hline \hline
On-source & On-source & Expected & Event & Line source limit & Diffuse limit & Gaussian limit \\
region & events & background & upper limit & & & \\
\hline \hline
$\pm2.0^\circ$    &   128    &   129.4   &  19.8 & $6.3 \times 10^{-5}$ &  
$6.6 \times 10^{-4}$ &   --  \\ 
\hline
$\pm4.4^\circ$    &   272    &   283.3   & 20.0 & -- & 
-- &  $4.8\times 10^{-4}$ \\
\hline
\end{tabular}
\caption{\label{tbl_results} Preliminary limits on the
  $\nu_{\mu}+\bar{\nu}_{\mu}$ flux at Earth from the outer galactic plane,
  for an $E^{-2.7}$ spectrum (systematic errors included). 
  Units on the line source limit are $\mathrm{GeV}^{-1} \mathrm{cm}^{-2}
  \mathrm{s}^{-1} \mathrm{rad}^{-1}$; units on the other two limits are 
  $\mathrm{GeV}^{-1} \mathrm{cm}^{-2} \mathrm{s}^{-1} \mathrm{sr}^{-1}$.}
\end{center}
\end{table}


\section{Conclusions}

Comparing the limit for a Gaussian flux profile in table \ref{tbl_results}
with the model prediction in section \ref{Introduction}, we find that the
sensitivity of this analysis is approximately two orders of magnitude above
the predicted flux.  IceCube, the $\mathrm{km}^3$-scale successor to
AMANDA-II, will have a larger effective area and better angular resolution
\cite{Ahrens04}.  Only for the most optimistic case, in which the source
profile is nearly a line, will the increase in angular resolution allow a
significantly smaller on-source window.  For five years of data from the
complete detector, the total improvement in sensitivity is just over one
order of magnitude.  Other approaches may be more sensitive -- for example,
we can focus only on dense areas of the ISM, such as the Cygnus region.
Also, recent calculations by Candia suggest that IceCube may be sensitive
to the flux from the Galactic Center using cascades from down-going
neutrinos \cite{Candia05}.  Detection of specific sources within the plane
may well precede discovery of a truly diffuse flux from the galactic disk.


\end{document}